\documentclass[letterpaper]{article} 
\usepackage[preprint]{aaai2027}  
\usepackage[hyphens]{url}  
\usepackage{graphicx} 
\usepackage{natbib}  
\usepackage{caption} 
\usepackage{algorithm}
\usepackage{algorithmic}

\usepackage{newfloat}
\usepackage{listings}
\DeclareCaptionStyle{ruled}{labelfont=normalfont,labelsep=colon,strut=off} 
\floatstyle{ruled}
\newfloat{listing}{tb}{lst}{}
\floatname{listing}{Listing}

\usepackage{booktabs}

\usepackage{caption}
\usepackage{amsthm}
\usepackage{multirow}

\usepackage{fontawesome5}   
\newcommand{\hficon}{\raisebox{-0.15em}{\includegraphics[height=1em]{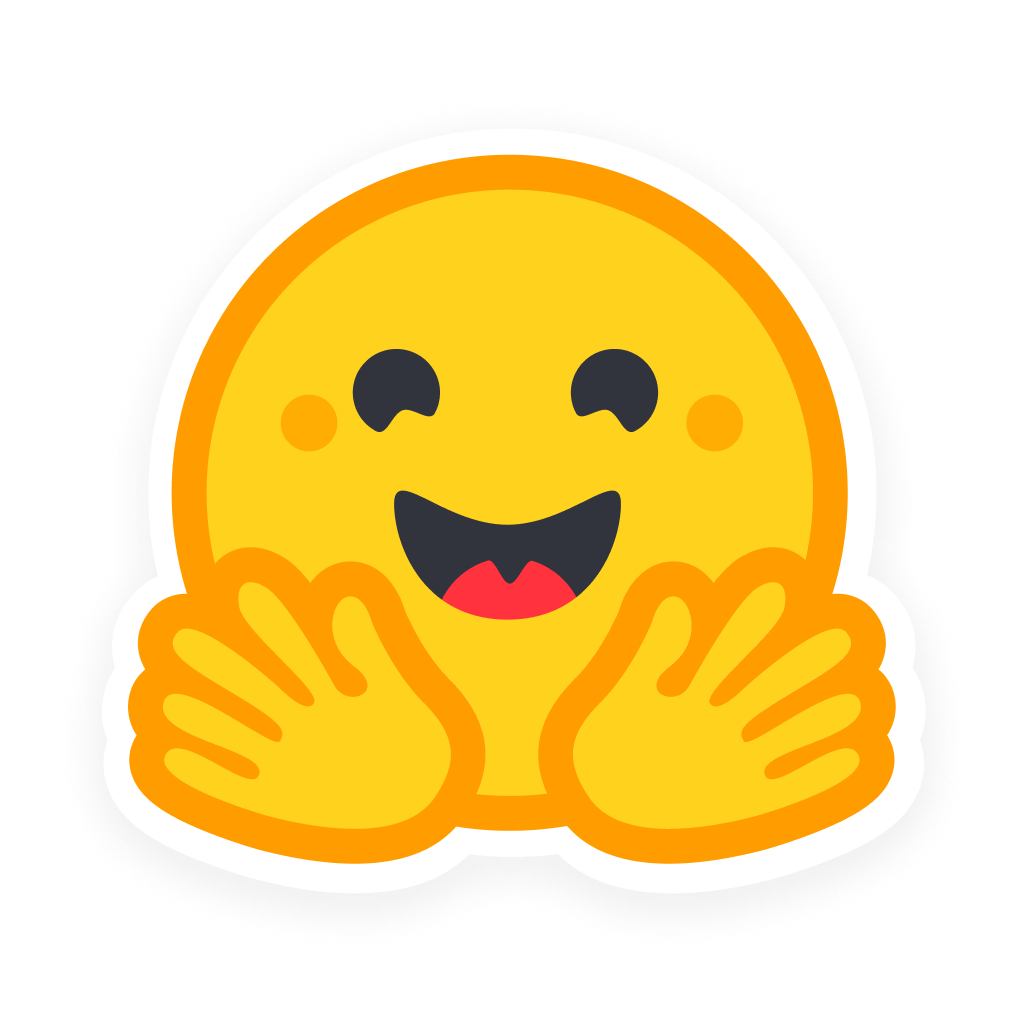}}}

\usepackage{subcaption}
\usepackage{graphicx}

\usepackage{amssymb}
\usepackage[mathscr]{euscript}

\usepackage{amsfonts,amsmath,amssymb}
\usepackage{xspace}

\usepackage{multicol}
\usepackage{enumitem}

\title{A Persona-based Rate Action Index}
\author{
    Hayden Helm\textsuperscript{\rm 1}\thanks{Corresponding author: \texttt{<first-name>}@helivan.io.} \& Andrew Dassori\textsuperscript{\rm2}
}
\affiliations{
    \textsuperscript{\rm 1}Helivan \& \textsuperscript{\rm 2}Wavelength Capital \\  \faGithub~Code:~\url{github.com/helivan-research/fomc-personas} \\
    \hficon~Data:~\url{huggingface.co/datasets/helivan/fomc-personas}
    \\
    Live Index: federalreserve.ai

}

\begin{document}
\maketitle

\begin{abstract}
We propose an index for predicting the U.S.\ Federal Open Market Committee (FOMC) decision to hike/hold/cut the current federal funds target rate based on how a collection of personas responds to current market conditions.
To construct the index, we collected a new dataset consisting of nearly $25{,}000$ retrievable chunks from publicly available
data. We partition the data into per-member corpora and use each as the retrieval database of a
generative system we refer to throughout as a ``persona''.
We first evaluate the personas across two complementary components of likeness: identifiability and
detectability.
Each persona's behavior is highly attributable (average member-conditional recall is $ 8\times $ chance) and generated content is nearly indistinguishable from held-out real content
($\hat\tau_{\mathrm{det}} = 0.23$ against a $0.15$ floor).
We then present evidence that query-conditioned representations of the personas capture members' monetary-policy stance relative to a known hawk--dove
reputational ordering (Kendall's $\tau = 0.63$, $p < 0.001$), substantially outperforming
retrieval-only representations.
These representations vary with time
and current market conditions and form the basis of our proposed persona-based rate action index.
For the $2022$--$2025$ period the index tracks the rate cycle (Kendall's $\tau = 0.68$, $p <
10^{-6}$) and can be used to construct a simple classifier that predicts per-meeting outcomes at
non-trivial accuracy ($0.69$ versus a $0.47$ base rate).
Importantly, the index outperforms informative baselines and leads the federal funds target rate by
roughly three quarters.
As far as we are aware, our results are the first to demonstrate the ability to capture time-varying group behavior via a collection of digital personas.
\end{abstract}

\section{Introduction}
\label{sec:intro}
The U.S.\ federal funds target rate is among the most consequential levers in the global economy. The rate
is set by the Federal Open Market Committee (FOMC) at eight scheduled meetings throughout the year.
Each decision to hike/hold/cut the rate impacts asset prices, borrowing costs and
the path of inflation and employment. Market participants therefore devote considerable time and effort to anticipate the
committee's actions: expectations of the next move are priced continuously in federal funds futures, and
surprises relative to those expectations move asset prices sharply \citep{kuttner2001monetary}. To
date, quantitative forecasts of rate actions have been market-implied (i.e., read from futures),
rule-based \citep{taylor1993discretion}, or distilled from the tone of central-bank communication
\citep{lucca2009measuring,shah2023trillion}. These approaches aggregate over the committee and do not
attempt to model the committee members' heterogeneous dispositions
\citep{istrefi2019fedwatchers,bordo2023perceived}.

Recent work in adjacent literatures has demonstrated that modeling individuals from their past behavior is possible.
For example, in generative surveying a profile-conditioned model reproduces survey responses
\citep{argyle2023outofone,brand2023llm} and can recover subgroup
opinions \citep{santurkar2023whose}, replicate human-subject experiments
\citep{aher2023simulating,horton2023homosilicus,yeykelis2024personas}, and, given rich per-person
data, match the answers of thousands of named individuals \citep{park2024thousand}.
Closest to our setting, \citet{helm2025congress} build a persona of each member of the U.S.\ Congress and predict their voting behavior on legislation.
These results share a structure we reuse: a person is modeled by a generative model with access to their past behavior, and the model is then queried to predict a future action.
The FOMC is a favorable setting for this approach as it is a (relatively) fixed body of public figures, each with a
long attributed record.

Concretely, we build a persona -- which, for our purposes, is a base model combined with a
member-specific retrieval database -- for each member from their public record and use
their behavior with respect to a collection of market-aware queries to construct an index for rate
action. Our contributions are threefold. \textbf{(i)~Data:} we present a new speaker-attributed
temporal corpus of FOMC members' public statements designed for retrieval-augmented generation. The data covers seventeen of nineteen sitting members
and includes $24{,}333$ chunks from $2006$ to $2026$ (\S\ref{sec:data},
Appendix~\ref{app:data}). \textbf{(ii)~A two-axis likeness check:} we verify each persona is faithful to
its member along two axes, \emph{identifiability} (between-persona) and \emph{detectability}
(within-persona) (\S\ref{sec:likeness}). \textbf{(iii)~A persona-based index (PBI):} comparing persona behavior across time, we find
that query-conditioned behavioral representations of the personas match a reputational ordering of the members (\S\ref{sec:stance}).
Importantly, the behavioral representations vary with market conditions and can be aggregated into a forward committee-level index. We validate the index against realized FOMC
decisions and relevant baselines (\S\ref{sec:index}).
As far as we are aware this is the first validation of using personas to track a time-varying covariate.


\subsection{Related Work}
\label{sec:related}
\paragraph{Synthetic respondents and digital twins.}
Beyond the generative-surveying results that motivate our approach (\S\ref{sec:intro}), a broad
literature conditions language models on personas to role-play identities or stand in for human
respondents \citep{chen2024persona,tseng2024persona,shanahan2023roleplay} and runs multi-agent
simulations in place of human social studies \citep{park2023generative,mcguinness2024mirroring}. The
promise is contested -- synthetic respondents can misstate distributions and hinge on prompt phrasing
\citep{dillion2023can,helm2026generative} -- which is why our analysis is grounded in per-member
detectability and out-of-sample validation. A more rigorous instance is the \emph{digital twin}: a
continuously updated computational counterpart of a system that supports valuable inference about it
\citep{nasem2024digitaltwins}. A growing line of work builds language-model twins of individuals and
groups -- benchmarks for persona-based behavior simulation \citep{li2025digitaltwins}, datasets
pairing thousands of people with their twins \citep{toubia2025twin,park2024thousand}, and
population-scale twins for simulating response to policy \citep{koaik2026social}. We build on the
framework of \citet{helm2025congress}, which sets requirements for a collection of personas to count
as a digital twin -- relevant and continuously updated data, high likeness, valid inference, and
useful predictions -- and instantiate one for the entire FOMC. We evaluate our personas against these
principles.
We further extend evidence of possible valid inference from a static quantity (e.g., a discrete vote on
legislation) to a dynamic one (e.g., continuously moving policy instrument), which requires conditioning the personas on a
time-varying environment not necessary in previous work.

\paragraph{Measuring FOMC stance and action.}
A rich literature extracts member disposition (often framed as the ``hawk (hike)--dove (cut) axis'') from
central-bank text: dictionary and lexicon scores
\citep{lucca2009measuring,tobback2017hawks,picault2017words}, computational-linguistics analyses of
transcripts \citep{hansen2018transparency,hansen2016shocking}, objective-function estimates
\citep{shapiro2022fedword}, vocal tone \citep{gorodnichenko2023voice}, and market-implied surprises
\citep{kuttner2001monetary}. Recent work applies language models to Fed communication directly,
classifying statement stance \citep{hansen2023chatgpt}, quantifying member dissent
\citep{peskoff2024gpt}, and building high-frequency sentiment indices from news coverage of Fed
communications \citep{wong2023fedspeak}, alongside datasets that label FOMC text for stance and
market impact \citep{shah2023trillion}. Most of this work scores statement- or institution-level
communication.
We instead
model each member as a generative persona and aggregate the members' inferred stances into a
committee index validated against realized rate decisions.


\section{Setup and Notation}
\label{sec:setup}
For our purposes, a generative system $f:\mathcal{Q}\to\mathcal{X}$ is a (random) mapping from a collection of queries to a response space.
A system's response $f(q)$ is assumed an independent sample from a query-conditioned distribution.
Content produced by a member of the FOMC is modeled as an observation from the query-conditioned distribution of a member-specific ground truth system.
In particular, let $\mathcal{F}^{*(t)}=\{f_1^{*(t)},\dots,f_n^{*(t)}\}$ be the collection of $n$ members at time $t$, where member $i$ is a system $f_i^{*(t)}$ and their statements are realizations of $f_i^{*(t)}(q)$ for a given $ q $.
We approximate each member-specific system with a \emph{persona} $f_i^{(t)}$, parameterized by a base model $f_{\mathrm{base}}$ and a member-specific retrieval database $D_i^{(t)}$.
The personas form the
collection $\mathcal{F}^{(t)}=\{f_1^{(t)},\dots,f_n^{(t)}\}$.
Here $D_i^{(t)}$ denotes member $i$'s statements dated $\le t$ (\S\ref{sec:data}) in the full $D_i$.
We let $g:\mathcal{X}\to\mathbb{R}^{p}$ denote a fixed embedding function that maps each response to a
$p$-dimensional vector.

In our setting, each member also has a time-varying system-level covariate
$y_i^{(t)}\in\mathbb{R}^{d}$ that we wish to recover -- such as their disposition to hike/hold/cut the
federal funds target rate given market conditions at time $t$.

\paragraph{Likeness.}
The first goal of this paper is to establish that, for the $D_i$ described in \S\ref{sec:data}, a
persona approximates its member: $f_i^{(t)}\approx f_i^{*(t)}$. We decompose this
\emph{likeness} into between- and within-persona components in
\S\ref{sec:likeness}.

\paragraph{System-level covariate prediction.}
The second goal is to recover each member's hike/cut/hold disposition from its persona across variable market conditions.

\section{Data and Corpus Construction}
\label{sec:data}
The FOMC comprises nineteen individuals: seven Board governors and twelve regional Reserve Bank
presidents. To construct each persona's member-specific corpus, we
draw on four document types: i)~speeches, ii)~testimony, iii)~FOMC meeting
transcripts (exact speaker labels, released on a five-year lag), and iv)~FOMC press conferences
(chairs only, no lag).

Given a document, we extract retrievable chunks with an LLM under a constrained schema. Each chunk
is a record $\langle$source, member id, time, is\_voting, is\_chair, topic, stance, embedding$\rangle$.
The topic is a free-text label that we classify into one of six themes: inflation and prices,
employment and labor, rates and policy, financial stability, growth and outlook, or other. The
stance is a self-contained sentence stating the member's position, and
its embedding is a $1024$-dimensional vector (\texttt{text-embedding-3-large},
\citealp{openai2024embeddings}). For example, one press-conference chunk has stance ``We want
to ensure that inflation remains near our symmetric 2 percent longer-run goal on a sustained
basis,'' topic ``inflation target,'' and theme ``inflation and prices.'' The booleans is\_voting
and is\_chair indicate whether the speaker
was a voting member and whether they were the chair on the statement's date, derived from the member
id and time against the FOMC voting rotation and chair tenure (null when a statement is undated).
We only admit a chunk to a member's corpus if their attribution is unambiguous.

\begin{figure}[t]
\centering
\includegraphics[width=\columnwidth]{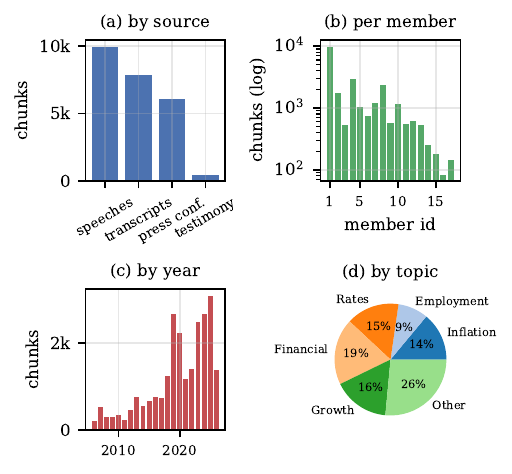}
\caption{Corpus composition ($24{,}333$ chunks). (a)~by source; (b)~per member, by member id
(Table~\ref{tab:roster}); (c)~by year of the source document; (d)~by topic theme. The core macro themes account for $\sim$three quarters
of the corpus.}
\label{fig:data}
\end{figure}

We onboard seventeen of the nineteen members. Two we cannot: one whose only outlet blocks automated
access at the network level, and one an interim appointee with no published record. The resulting
corpus spans $883$ documents and $24{,}333$ chunks.
Per-member counts range from $285$ documents ($9{,}855$ chunks) for the chair to $6$ documents
($83$ chunks) for the most recently appointed president.
Coverage is continuous from 2006 to the present.
We also create a biography for each member to further ground the persona.
Biographies share a common format and are used as the persona's system prompt.
Figure~\ref{fig:data}
summarizes the corpus. Appendix~\ref{app:data} documents every source, the constrained extraction
schema and chunking, and the per-stage filters. Appendix~\ref{app:roster} provides per-member
corpus size, channels, and collection start dates.

Ours is not the first dataset consisting of public FOMC data.
Existing
datasets annotate text at the sentence or statement level for classification or measurement \citep{lucca2009measuring,hansen2018transparency,gorodnichenko2023voice,shah2023trillion, kanganis2025opfed}.
None attribute text to individual members across multiple sources and none include pre-computed embeddings for retrieval-based personas.
We provide a more in-depth comparison of our dataset to others in Appendix~\ref{app:data} in Table~\ref{tab:datasets}.

\section{Likeness}
\label{sec:likeness}
\begin{figure*}[t!]
\centering
\includegraphics[width=0.9\textwidth]{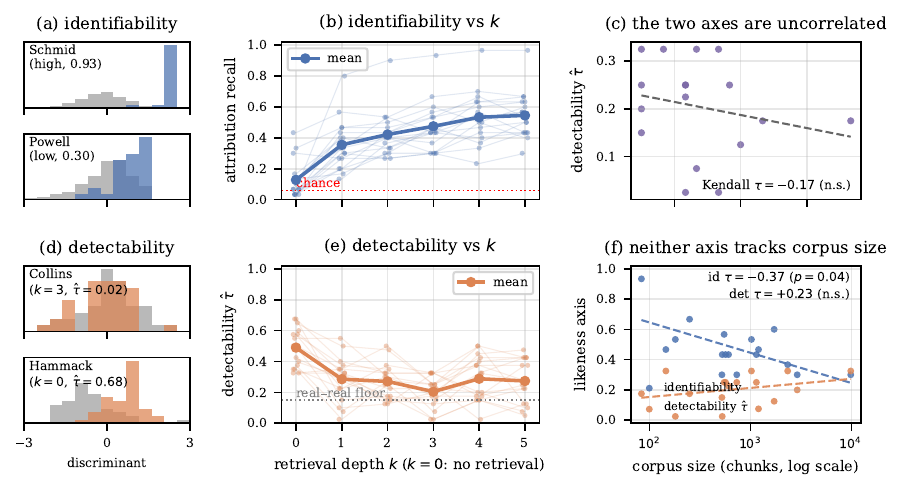}
\caption{Likeness of the persona collection, measured along two axes:
\emph{identifiability} (top row, blue) and \emph{detectability} (bottom row, orange).
(a)~Identifiability example: the $17$-way attribution classifier's log-probability that a response is
the named member, on that member's own responses vs.\ all others' (grey), for a high- (Schmid) and a
low-identifiability (Powell) member. (b)~Per-member identifiability (attribution recall) vs.\
retrieval depth $k$ (faint~=~members, bold~=~mean; chance $=1/17$). (d)~Detectability example:
generated continuations vs.\ the member's real ones (grey) for a low-detectability (Collins, $k{=}3$)
and a high-detectability (Hammack, $k{=}0$) member. (e)~Per-member $\hat\tau_{\mathrm{det}}$ vs.\
$k$ (real--real floor in grey). (c)~Per-member identifiability vs.\ detectability ($k{=}3$).
(f)~Both axes ($k{=}3$) vs.\ per-member corpus size (log scale).}
\label{fig:likeness}
\end{figure*}
We build up to the persona-based action index by first demonstrating that the collected corpora parameterize personas that have high likeness to their corresponding member.
Unlike previous work, we break this down into two complementary components: (i)~\emph{identifiability}
and (ii)~\emph{detectability}. We formalize each as a property of a query set and the risk of a
classification problem.

Let $Q=\{q_1,\dots,q_m\}$ be a collection of queries and, for a system $f$, let $P_f(q)$ be the
distribution of the response $f(q)$ and $P_f(Q)$ the distribution of its responses over $Q$. We let
$h\colon\mathcal{X}\to\mathcal{Y}$ be a decision function on $\mathcal{X}$ and $\ell$ an appropriately
defined loss. We define $R(h):=\mathbb{E}_{(X,Y)}[\ell(h(X),Y)]$ as the risk of $h$ and
$R_i(h):=\mathbb{E}_{X\mid Y=i}[\ell(h(X),i)]$ its class-conditional risk. Given
a fixed set of decision functions $\mathcal{H}$, we let $h^{*}=\arg\min_{h\in\mathcal{H}}R(h)$ be the
decision function that achieves minimal risk and $h^{c}$ a chance-level classifier. For simplicity we
assume zero-one loss.

The two components of likeness can then be described as properties of two classification problems:
\begin{enumerate}
\item[(i)] \emph{identifiability} is the $n$-way problem of attributing a response $X\sim P_{f_i}(Q)$
to its persona, assuming uniform class-conditional priors. All personas are identifiable if the best
classifier in $\mathcal{H}$ outperforms $h^{c}$ on every persona:
\[
  \max_{i}R_i(h^{*}_{\mathrm{id}})\;<\;R_i(h^{c}_{\mathrm{id}}).
\]
Estimating identifiability requires each persona to respond to the same set of queries.
\item[(ii)] \emph{detectability} is the binary problem of separating a member's real responses
$P_{f_i^{*}}(Q)$ from its persona's generated ones $P_{f_i}(Q)$; following \citet{helm2025congress},
we require low detectability:
\[
  \tau_{\mathrm{det}}\;=\;1-R(h^{*}_{\mathrm{det}})/R(h^{c}_{\mathrm{det}})\;\approx\;0
\]
($\tau_{\mathrm{det}}{=}0$ indistinguishable, $\tau_{\mathrm{det}}{=}1$ always detected).
Estimating detectability requires each persona to produce content paired with content known to be
written by its member.
\end{enumerate}

\paragraph{Personas.}
We use \texttt{gpt-4o-mini} \citep{openai2024gpt4} as $f_{\mathrm{base}}$.
Each persona has a member-specific biography as
its system prompt and access to $D_i$.
We embed all text with \texttt{text-embedding-3-large} \citep{openai2024embeddings}. Given a query $q$, we score every stance $s\in D_i$ by the cosine similarity between its embedding and the query's and place the top $k$ in context, for $k\in\{0,\dots,5\}$ ($k{=}0$ is the biography-only persona).

\paragraph{Methods.}
For \emph{(i) identifiability}, every persona answers the same query set $Q$ -- $30$
monetary-policy questions designed to elicit member behavior.
We embed each response and train a linear classifier to recover which persona produced it.
A persona is identifiable when its
attribution recall exceeds chance ($1/17 = 0.059$).
For \emph{(ii) detectability}, we use the seeded completions proposed by \citet{helm2025congress}: for
each member we hold out $40$ of their real stances ($\ge 8$ words), reveal the opening words of each,
and have the persona complete it. We train a linear classifier to predict the source of the completed
stance and report detectability ($\hat\tau_{\mathrm{det}}$). As a reference we also report the average
\emph{real--real} detectability -- the detectability between two independent samples of a member's own
real content. It is a lower bound on the average persona detectability: no persona can
resemble a member more closely than the member's own writing does.
The first column of Figure~\ref{fig:likeness} shows high- and low-scoring examples on each axis.

\paragraph{Results.}
The center columns (b,~e) of Figure~\ref{fig:likeness} show how identifiability and detectability vary
with the retrieval depth $k$. The biography-only persona ($k{=}0$) is weakly identifiable -- mean recall $0.13$ ($2.2\times$ chance),
with $5$ of $17$ members below chance -- and highly detectable ($\hat\tau_{\mathrm{det}}\approx0.5$).
By $k{=}3$ every persona has enough member-specific signal (mean recall $0.48$, minimum $0.30$; average $\hat\tau_{\mathrm{det}}=0.23$ relative to average real--real lower bound of $ 0.15 $).

The top-right panel (c) shows that the two components are uncorrelated across members (Kendall
$\tau{=}{-}0.17$, n.s.) -- an identifiable persona may be detectable and vice versa.
The bottom-right panel (f) plots both axes against corpus size and shows that neither is a
data-volume artifact. Identifiability is, if anything, \emph{negatively} associated with corpus size
(Kendall $\tau{=}{-}0.37$, $p{=}0.04$): the smallest twin (Schmid, $83$ chunks) is the most
identifiable ($0.93$).
This is likely a property of the FOMC since governors and the chair have the largest corpora but are usually tightly aligned in their stances.

\section{Stance Prediction}
\label{sec:stance}
\begin{figure*}[t!]
\centering
\includegraphics[width=0.9\textwidth]{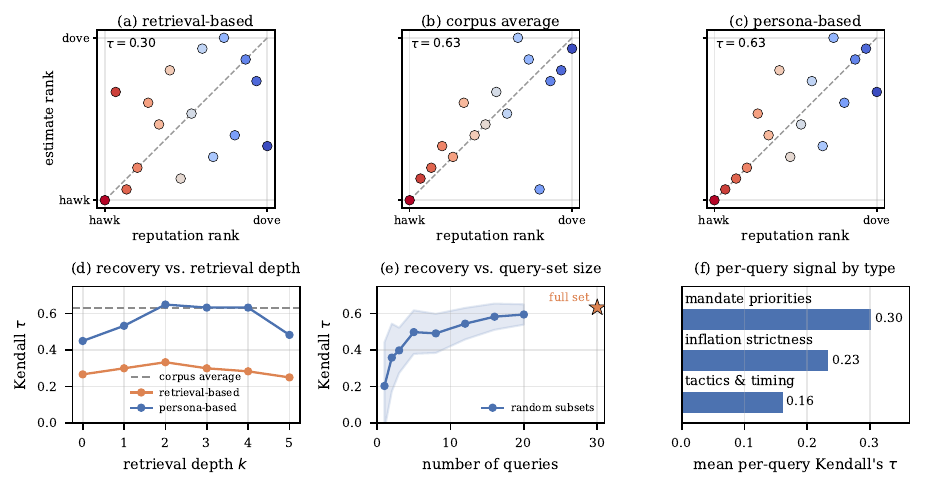}
\caption{Stance recovery against an external hawk--dove reputational ordering of the $16$ scored
members. (a--c)~At $k{=}3$, each panel ranks the members by their reputation (horizontal) against the
rank induced by one estimate (vertical); the dashed diagonal is perfect recovery (Kendall $\tau$ in
the title, colour~=~reputation): (a)~retrieval-based, (b)~corpus average, (c)~persona-based.
(d)~Per-estimate $\tau$ vs.\ retrieval depth $k$ (corpus average has no retrieval, dashed; $k{=}0$ is
the biography alone). (e)~Persona-based $\tau$ vs.\ query set size. (f)~Mean per-query Kendall's $\tau$ by
query type.}
\label{fig:stance}
\end{figure*}
We have established that the personas behave like their corresponding member. We now use them to
predict the members' relative position on the hawk--dove spectrum by their responses to a collection
of queries. We use a news-derived ordering of the members along the hawk--dove spectrum
\citep{istrefi2019fedwatchers,bordo2023perceived} to validate and compare different ways of
estimating the stance ordering.

\paragraph{Stance representations.}
We compare three representations, each providing a single vector per member summarizing their position:
\begin{enumerate}
\item[(i)] \emph{corpus average}: the mean of the embeddings of every statement in $D_{i}$;
\item[(ii)] \emph{retrieval-based}: the mean embedding of the top-$k$ statements
retrieved from $D_{i}$, averaged over a query set $Q$;
\item[(iii)] \emph{persona-based}: the mean embedding of the response the persona
generates, averaged over a query set $Q$.
\end{enumerate}
Given the per-member representations, we project them onto a spectrum defined by the mean embeddings of
canonically hawkish and dovish statements, $\widehat{\mathrm{hawk}}$ and $\widehat{\mathrm{dove}}$.
In particular, letting $u=\widehat{\mathrm{hawk}}-\widehat{\mathrm{dove}}$, we predict each member's stance with
$\widehat{y}_i=\langle\psi_i(Q),u\rangle$.
We report the retrieval-based and persona-based estimates across retrieval depths $k$,
evaluating each projection by Kendall's $\tau$ against the reputational ordering. The curated set of
$m=30$ monetary-policy questions, together with the hawkish and dovish anchor statements, is provided
in Appendix~\ref{app:stance}.

\paragraph{Results.}
The results for recovering the reputational hawk--dove ranking are provided in Figure~\ref{fig:stance}.
The top row shows the estimated ranking of the $\widehat{y}_{i}$ for each per-member representation at
$k=3$. Panel (d) shows the sensitivity of recovery to retrieval depth. The retrieval-based estimate is
weak at every retrieval depth (Kendall $\tau=0.23$--$0.32$, not significant). The corpus average
recovers the reputational ordering well ($\tau=0.63$, $p<0.001$), and the persona-based estimate
matches or exceeds it once a few statements are retrieved -- peaking at $k{=}2$ ($\tau=0.65$) and
holding $\tau=0.63$ through $k{=}4$ (Figure~\ref{fig:stance}d). The corpus--reputation agreement is
expected -- the reputation is a tenure average, which any career-level data-driven summary will
match almost by construction.

\begin{figure*}[t!]
\centering
\includegraphics[width=0.9\textwidth]{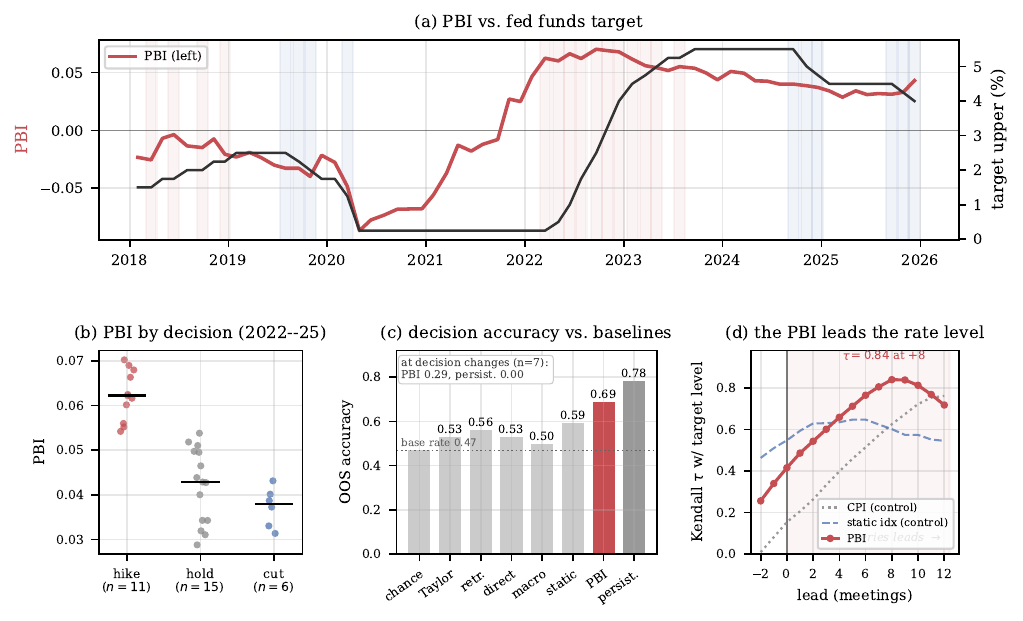}
\caption{The persona-based index (PBI). (a)~The PBI (red, left axis) moderately tracks the
fed funds target (black, right axis) over 2018--2025. (b)~PBI by realized decision
(2022--25): hikes separate cleanly (hike-vs-rest AUC $1.00$), while restrictive-era cuts and holds
partially overlap. (c)~Walk-forward decision accuracy (2022--25): the PBI ($0.69$) beats chance, the Taylor rule, a fitted
macro model, and the retrieved-text and static-query controls; only persistence is higher. (d)~Kendall's $\tau$ between each series and the funds-rate level as
the series is slid forward. The PBI's correlation rises from $0.42$ (contemporaneous) to $0.84$ at a
lead of $+8$ meetings ($\approx$3~quarters), significantly exceeding both controls.}
\label{fig:index}
\end{figure*}
The last two panels of Figure~\ref{fig:stance} show the effect of the query-set size (panel~e) and
type (panel~f) on the quality of the persona-based estimates. For panel (e) we randomly sample query
sets of sizes $\{1,\dots,20\}$ from the $30$ curated queries to estimate the reputational ordering; we
note diminishing returns in recovery beyond $m\approx10$. For panel (f) we instead score a separate,
balanced pool of $72$ queries spanning $12$ hawk--dove facets, grouped into three topics (mandate
priorities, inflation strictness, and tactics \& timing). Ranking members from each single query
($m{=}1$) and averaging Kendall's $\tau$ within each topic, questions about a member's mandate
priorities discriminate most ($\tau=0.30$) while timing- and tactics-oriented questions carry the
least signal ($\tau=0.16$). We compare the persona-based ranking generated with \texttt{gpt-4o-mini}
to those generated with \texttt{ministral-8b} in Appendix~\ref{app:generator}; they have large
agreement (Kendall's $\tau=0.69$).

\section{A Rate-Action Index}
\label{sec:index}
Reputational stance estimation validates basic inference with the personas on a static target. We now
extend stance prediction into a rate-action index that responds to market conditions, by adding
time-varying information to each query and tracking the personas' behavior over time.

\paragraph{Incorporating market conditions.}
The FOMC makes its decision to hike/hold/cut the federal funds rate based on several
factors, including headline and core inflation, the unemployment rate, and the current
target range for the rate, all available from the FRED repository of economic data \citep{fred}. We
summarize current market conditions with a single string $c^{(t)}$:
\begin{quote}\itshape\small
Current U.S.\ economic conditions as of \{\texttt{date}\}: CPI inflation is running at about
\{\texttt{cpi}\} year over year; core PCE inflation is about \{\texttt{core\_pce}\} year over year (the
FOMC targets 2\%); the unemployment rate is \{\texttt{unemployment}\}; the current federal funds target
range tops out at \{\texttt{target}\}.
\end{quote}
We prepend $c^{(t)}$ to every query, forming $Q^{(t)}=\{c^{(t)}{+}q_1,\dots,c^{(t)}{+}q_m\}$.

Given $Q^{(t)}$ we generate stance representations for each member using $D_i^{(t)}$, the member's
content publicly available strictly before $t$. FOMC meeting transcripts (32\% of the corpus) are
dated at the meeting they record but published only after a $\approx$5-year embargo, so filtering on
the statement date alone would let the backtest retrieve text that was not public at time $t$; we
date transcripts at their release and apply a strict day-before cutoff, so nothing a persona reads
postdates -- or was unpublished at -- the meeting being predicted. We project each representation
onto the fixed hawk--dove axis $u$ (\S\ref{sec:stance}) and average across members to produce a
committee-level, persona-based index (PBI).

\paragraph{Validation.}
We validate PBI on two tasks: i) \emph{tracking} the federal funds target rate and ii) \emph{predicting} the
hike/hold/cut outcome of each meeting. For tracking, we measure how well PBI's level co-moves with the
per-meeting basis-point move over the $64$ FOMC meetings from 2018 to 2025, measured by Kendall's
$\tau$. For prediction, we fit a linear hike/hold/cut classifier on the committee index and evaluate
it walk-forward. We report three different accuracies: 3-class classification, move versus hold, and
hike versus rest.

We compare PBI to seven baselines under a walk-forward protocol. Three are ablations to localize the signal:
\begin{enumerate}\setlength{\itemsep}{0pt}
\item \emph{Macro}: a classifier fitted on the same macroeconomic inputs (inflation, unemployment, the target
level, and their momentum) -- does the persona signal add over the raw macro data?
\item \emph{Retrieved-text}: a no-generation index that projects each member's date-$\le t$ retrieved
chunks onto the same axis -- does generation beat scoring the recent text?
\item \emph{Static query set}: the same generation pipeline over the fixed queries, \emph{without} the
time-varying briefing $c^{(t)}$ -- does conditioning on the economy beat a static query set?
\end{enumerate}
The other four are simple baselines: \emph{chance} (always predict hold), \emph{persistence}
(repeat the previous meeting's decision), a fixed \emph{Taylor rule} \citep{taylor1993discretion}, and a \emph{direct vote} (ask
each persona to vote hike/hold/cut directly).

\begin{table}[t]
\centering
\small
\begin{tabular}{@{}lcccc@{}}
\toprule
 & Track & \multicolumn{3}{c}{Classification (2022--25)} \\
\cmidrule(lr){2-2}\cmidrule(lr){3-5}
Method & $\tau$ & 3-class & hike/rest & move/rest \\
\midrule
\multicolumn{5}{@{}l}{\textit{Baselines}}\\
Chance (hold)             & ---     & 0.47 & 0.66 & 0.47 \\
Persistence               & 0.76    & 0.78 & 0.88 & 0.78 \\
Taylor rule               & 0.57    & 0.53 & 0.72 & 0.53 \\
Direct vote               & 0.60    & 0.53 & 0.75 & 0.53 \\
\midrule
\multicolumn{5}{@{}l}{\textit{Ablations}}\\
Macro (fitted)            & $0.49^{\dagger}$ & 0.50 & 0.69 & 0.50 \\
Retrieved text            & $-0.08$ & 0.56 & 0.69 & 0.62 \\
Static query set          & $-0.05$ & 0.59 & 0.78 & 0.62 \\
\midrule
\textbf{PBI (all)}        & 0.68    & 0.69 & 0.88 & 0.69 \\
PBI (voting only)         & 0.70    & 0.72 & 0.91 & 0.72 \\
PBI (chair only)          & 0.68    & 0.69 & 0.88 & 0.69 \\
\bottomrule
\end{tabular}
\caption{All methods on both validations over the 2022--25 window. Tracking is Kendall's $\tau$
relative to the realized per-meeting move; the three walk-forward classification accuracies are
3-class hike/hold/cut plus \emph{hike/rest} and \emph{move/rest} one-vs-rest detection;
$^{\dagger}$the fitted macro model has no unsupervised level, so its $\tau$ uses its out-of-sample
predicted move.}
\label{tab:baselines}
\end{table}

\paragraph{Results.}
The PBI traces the cycle (Figure~\ref{fig:index}a): mildly hawkish in 2018, dovish through the
2019--21 easing, climbing ahead of the 2022 pivot, and peaking in 2022--23 before moderating through
the 2024--25 easing to a level still elevated relative to the pre-tightening cycle.
Its level co-moves with the realized move (Kendall's $\tau=0.35$ over the full period; $\tau=0.68$,
$p<10^{-6}$, over $2022$--$2025$) and only persistence tracks comparably (Table~\ref{tab:baselines}).

The PBI classifies hike/hold/cut at $0.69$ against a $0.47$ base rate, with strong hike detection (hike/rest $0.88$; hike-vs-rest AUC on
the raw level $1.00$). It beats chance, the Taylor rule ($0.53$), and the retrieved-text index
($0.56$).
Most pointedly, it beats a macro classifier fitted on the same briefing variables
$c^{(t)}$ the personas have access to ($0.50$), and removing the briefing entirely (static query
set) collapses tracking to $\tau\approx0$. Generation alone is not enough, though --
a direct hike/hold/cut vote collapses toward the hold base rate ($0.53$) -- so the predictive power
sits in both the generation and its projection onto the stance axis.

The PBI does not beat persistence on raw accuracy ($0.69$ vs.\ $0.78$), though persistence is wrong
at every pivot out of a regime by construction. At the seven decision changes in the evaluation
window it scores $0.00$ and the PBI $0.29$.
The PBI's residual deficit is concentrated entirely in cuts ($0/6$).
The PBI conditioned on the meeting's decision is shown in Figure~\ref{fig:index}b: the level
separates cuts from holds only partially, though not enough for the walk-forward classifier to ever call a cut.

\paragraph{The index leads the rate level.}
Sliding the index forward $k$ meetings \emph{raises} its rank correlation
with the funds-rate level, from $\tau=0.42$ contemporaneously to $\tau=0.84$ at $k{=}8$
($\approx$3~quarters; see Figure~\ref{fig:index}d).
The lead is significant (block-bootstrap 95\% CI $[0.15, 0.89]$), and the PBI leads the realized
rate better than a CPI-only baseline (paired difference 95\% CI $[0.05, 0.49]$).
That is, the members' aggregated public language -- as picked up by the personas -- turns hawkish or dovish roughly three quarters before the policy rate arrives at the corresponding level.

\section{Discussion}
\label{sec:discussion}
We built a speaker-attributed, temporally-resolved corpus of FOMC communications ($24{,}333$ chunks,
$2006$--$2026$, seventeen of nineteen sitting members) and used it to construct retrieval-augmented
personas of the committee. The personas pass both likeness checks -- \emph{identifiability} between
personas and \emph{detectability} within each -- and, when queried, recover an external reputational
hawk--dove ordering ($\tau=0.63$). Conditioned on each meeting's market environment and aggregated
over the committee, they form the persona-based index (PBI), which -- under a strict point-in-time
corpus -- reconstructs the $2018$--$2025$ rate cycle ($\tau=0.68$ out-of-sample), classifies
per-meeting hike/hold/cut at $0.69$ against a $0.47$ base rate, and, most distinctively, \emph{leads}
the funds-rate level by roughly three quarters ($\tau$ rising $0.42{\to}0.84$ when slid forward).
As far as we are aware this is the first quantitative demonstration that the outputs of a collection of personas can track a time-varying system-level covariate.

\paragraph{Limitations.}
Several properties of the setting bound these results. Committee members share an industry-specific
rhetoric, have overlapping backgrounds, and rarely dissent formally, and per-member corpora span $83$
to $9{,}855$ chunks -- all of which could limit identifiability and detectability.
The personas
nonetheless demonstrate high likeness on both axes.

The corpora draw exclusively on official public
sources (speeches, testimony, meeting transcripts, and, for chairs, press conferences). This keeps
document types uniform but omits potentially informative material such as media interviews, social
media, and prior academic writing.

The evaluation window is also temporally bounded. The committee operated under a single chair
(Jerome Powell, February 2018 through May 2026) for the entire backtest, and the out-of-sample window
contains only $32$ meetings; communication norms under new leadership may shift how well personas
built on this record represent their members. Separately, the base model's training data ends in late
2023, so early ``predictions'' could partially reflect information already in the weights; on the
post-cutoff meetings alone the classifier matches but does not beat the hold-heavy base rate, while
the lead profile persists. Two further robustness notes: with embargoed transcripts naively dated at
the meeting (i.e., no point-in-time filter) accuracy inflates to $0.75$, and the committee mean is
robust to the choice of aggregation (Appendix~\ref{app:methods}).

Acknowledging these limitations, the synthesis of persona-based and economic modeling in the PBI
both tracks the aggregate policy decisions of the FOMC and holds predictive value -- the index
regularly leads the federal funds target rate (Figure~\ref{fig:index}a,\,d) -- for decisions that
are among the most scrutinized in the global economy.

\subsubsection*{Future Work}
\label{sec:future}
Three directions follow directly. (i)~The PBI's honest boundary is cut-versus-hold: a recalibration
cut and a hold-at-restrictive are stance-equivalent on a single hawk--dove axis, so a second,
``easing-urgency'' dimension learned the same way could separate the easing turns the single axis
collapses. (ii)~The briefing $c^{(t)}$ is a point-in-time snapshot. Conditioning on the recent macro
trajectory could supply the directional signal a level alone withholds. (iii)~Because the
twins are the present committee, the contributing roster thins historically (6 of $\sim$12 voters in
2018, rising to 16 by 2025); twins of former members, built from the same pipeline, would extend the
backtest.

\paragraph{Reproducibility statement.}
All scripts for the data collection and experiments are public: code at \url{github.com/helivan-research/fomc-personas},
and the corpus (with precomputed embeddings) at
\url{huggingface.co/datasets/helivan/fomc-personas}. A live version of the index, updated as new
statements are published, is served at \url{federalreserve.ai}.

\clearpage

\bibliography{references}

\clearpage
\appendix
\onecolumn

\section{Data Collection, Chunking, and Filtering}
\label{app:data}
This appendix documents the corpus-construction pipeline end to end: the public sources, how
documents are discovered, how raw text is segmented into passages, how each passage is
distilled into retrievable chunks, and the filters applied at each stage. All sources are public; no
authentication, paywalled, or private data is used. Every chunk carries the URL and access
timestamp of its source document.

\subsection{Sources and discovery}
A member's documents are gathered from the public outlets appropriate to their office
(Table~\ref{tab:sources}). Board governors publish through the Federal Reserve Board; regional
presidents publish through their home bank and, uniformly, through the Bank for International
Settlements (BIS) speech archive; all FOMC participants appear in the meeting
transcripts.

\begin{itemize}
\item \textbf{Board speeches and testimony.} The Board exposes machine-readable feeds
(\texttt{/json/ne-speeches.json}, \texttt{/json/ne-testimony.json}); we keep items whose speaker
field matches the member's name. Coverage begins at the member's first listed item and has no
embargo.
\item \textbf{FOMC meeting transcripts.} Verbatim transcripts are released on a five-year lag.
We enumerate meetings from the Board's per-year historical index pages (regex over
\texttt{FOMC\textit{YYYYMMDD}meeting.pdf}), download each meeting PDF, and segment it into
speaker-attributed turns. The transcripts in the corpus currently end in 2020 (the latest released
under the five-year embargo at collection time); this is the single largest constraint on recent
coverage.
\item \textbf{FOMC press conferences.} Chair-attributed press-conference transcripts carry no
embargo and are discovered from the same historical index (through 2020) and the FOMC calendar
page (2021$+$).
\item \textbf{Regional-president speeches via BIS.} The BIS archive (\texttt{bis.org/cbspeeches})
aggregates Federal Reserve regional speeches in one uniform place. We resolve a president's BIS
author page (by constructing candidate slugs, then falling back to paging the global index until a
speech by that speaker appears) and collect every speech linked there. Because BIS is a single
uniform source, it is the most robust president channel and the one used to onboard presidents
whose home-bank site refuses programmatic access.
\item \textbf{Regional-bank home pages.} For some presidents we additionally scrape the home bank's
speech repository. These are heterogeneous: discovery is by rendered listing page or by
\texttt{sitemap.xml}, with per-bank link and date patterns; bodies come from the HTML article or,
when too short, the linked PDF.
\item \textbf{Biographies.} A single official biography per member (Board or home-bank page,
normalized to a uniform register) grounds the persona's system prompt; it carries no time dimension
and is not part of the chunk corpus.
\end{itemize}

\begin{table}[h]
\centering
\small
\begin{tabular}{@{}lll@{}}
\toprule
Source & Applies to & Embargo / bound \\
\midrule
Board speeches \& testimony (JSON feeds) & Governors & none \\
FOMC meeting transcripts (PDF) & All participants & 5-year lag \\
FOMC press conferences (PDF) & Chairs & none \\
BIS speech archive & Regional presidents & none \\
Regional-bank speech pages & Some presidents & none \\
Official biography & All members & static \\
\bottomrule
\end{tabular}
\caption{Public sources per office. The five-year transcript embargo is the binding limit on recent
coverage; speeches and press conferences are contemporaneous.}
\label{tab:sources}
\end{table}

\subsection{Passage segmentation}
Each raw document is split into passages of at most $6{,}000$ characters by a hierarchical
splitter: paragraphs (double-newline boundaries) are accumulated up to the limit, and any single
paragraph exceeding the limit is further split on sentence boundaries. Transcripts are first
segmented into speaker turns and only the target member's turns are retained (turns shorter than
$80$ characters are dropped as procedural). This preserves argument-level context within a passage
while bounding the size handed to the extraction model.

\subsection{Chunk extraction}
Each passage is distilled into zero or more chunks by a structured-output LLM call
(\texttt{gpt-4o-mini}, $8$-way passage parallelism). The model is instructed to extract only the
speaker's own expressed views, ignoring procedural remarks and positions attributed to
others, and to return, for each chunk, a \texttt{topic} (a noun phrase naming the subject), a
\texttt{stance} (one self-contained sentence stating the position), and a verbatim \texttt{quote}
grounding it. The \texttt{stance} sentence becomes the chunk's \texttt{text} (the field that is
embedded and retrieved against), so retrieval and generation operate on normalized,
self-contained positions rather than raw prose. We then assign each chunk a coarse \texttt{theme}
(keyword bucketing of the \texttt{topic}) and, for FOMC members, the as-of-date \texttt{is\_voting}
and \texttt{is\_chair} roles derived from the member and statement date (\S3 of the main paper).

\subsection{Filtering}
Filters are applied at every stage. (i)~\emph{Speaker attribution}: Board items are kept only on
a name match in the speaker field; transcript turns only when attributed to the member; bank-page
bodies only when the surname appears in the article head or byline. (ii)~\emph{Length}: transcript
turns below $80$ characters and speech bodies below $400$ characters are discarded, and overly short
HTML bodies trigger a PDF fallback. (iii)~\emph{Date}: items are bounded below by a per-member
start date and, for transcripts, by the five-year embargo. (iv)~\emph{Deduplication}: when a
member's twin is updated, incoming chunks whose \texttt{text} already exists in the stored corpus
are dropped, so re-runs are idempotent and additive. Every Fed source carries
\texttt{probabilitySpeaker}~$=1.0$ (attribution is certain from speaker headers and verified pages);
the field is retained so the same schema supports lower-confidence sources elsewhere.

\subsection{Chunk schema}
Each chunk written to the per-member parquet has the fields:
\texttt{text} (the stance sentence, embedded and retrieved on),
\texttt{topic}, \texttt{theme} (one of six topic themes),
\texttt{quote} (verbatim support),
\texttt{stance} (equal to \texttt{text}),
\texttt{handle} (member name, the chunk's member id),
\texttt{postedAt} (date of the underlying document, the chunk's time),
\texttt{is\_voting}, \texttt{is\_chair} (as-of-date roles derived from the member and date; null when undated),
\texttt{source} (document type, e.g.\ \texttt{fed\_speech}, \texttt{fomc\_transcript}),
\texttt{sourceId}, \texttt{postUrl}, \texttt{accessedAt},
\texttt{probabilitySpeaker}, and
\texttt{embedding} ($1024$-d \texttt{text-embedding-3-large}, attached at persona-creation time).

\subsection{Relation to existing FOMC datasets}
Table~\ref{tab:datasets} summarizes how our corpus differs from prior FOMC text resources along the
dimensions discussed in \S3 of the main paper.

\begin{table}[h]
\centering
\footnotesize
\setlength{\tabcolsep}{4pt}
\begin{tabular}{@{}lcccc@{}}
\toprule
                              & Member & Multi- & Stance & Pre-comp. \\
Dataset                       & -level & source & labels & embed. \\
\midrule
Trillion Dollar Words         & $\times$     & \checkmark   & \checkmark   & $\times$     \\
Op-Fed                        & \checkmark   & $\times$     & \checkmark   & $\times$     \\
Hansen et al.                 & \checkmark   & $\times$     & $\times$     & $\times$     \\
Lucca \& Trebbi               & $\times$     & $\times$     & \checkmark   & $\times$     \\
Voice of Monetary Policy$^\dagger$ & \checkmark & $\times$   & $\times$     & $\times$     \\
\textbf{Ours}                 & \checkmark   & \checkmark   & \checkmark   & \checkmark   \\
\bottomrule
\end{tabular}
\caption{FOMC text resources compared along the dimensions our method requires.
\emph{Member-level}: text resolved to individual FOMC members. \emph{Multi-source}: two or more
document types. \emph{Stance labels}: carries stance/opinion annotation. \emph{Pre-comp.\ embed.}:
ships pre-computed embeddings for retrieval. $^\dagger$Chair only, vocal-tone modality. Citations
as in \S3 of the main paper.}
\label{tab:datasets}
\end{table}

\section{Member Roster}
\label{app:roster}
Table~\ref{tab:roster} lists the onboarded members, their office, the channels collected, and the
per-member collection start date. This start date is set to each member's earliest plausible
Federal Reserve activity: governors and long-serving presidents are collected from as early as
2006, while members appointed in 2018 or later start near their tenure (an earlier start yields no
additional chunks under their name, since the speaker filter admits only their own statements). Coverage is
therefore continuous in aggregate from 2006 to the present, with depth tracking tenure.

\begin{table}[h]
\centering
\small
\begin{tabular}{@{}rllllr@{}}
\toprule
ID & Member & Office & Channels & From & Chunks \\
\midrule
$1$  & Jerome H. Powell      & Chair (Board)            & Sp, Te, Tr, PC & 2012 & $9{,}855$ \\
$2$  & Kevin M. Warsh        & Governor                 & Sp, Te, Tr     & 2006 & $1{,}716$ \\
$3$  & Philip N. Jefferson   & Governor                 & Sp, Te, Tr     & 2018 & $521$ \\
$4$  & Michelle W. Bowman    & Governor                 & Sp, Te, Tr     & 2018 & $2{,}904$ \\
$5$  & Michael S. Barr       & Governor                 & Sp, Te, Tr     & 2018 & $1{,}022$ \\
$6$  & Lisa D. Cook          & Governor                 & Sp, Te, Tr     & 2018 & $731$ \\
$7$  & Christopher J. Waller & Governor                 & Sp, Te, Tr     & 2018 & $1{,}201$ \\
$8$  & John C. Williams      & President, New York      & Tr, BIS        & 2011 & $2{,}337$ \\
$9$  & Mary C. Daly          & President, San Francisco & Tr             & 2018 & $566$ \\
$10$ & Neel Kashkari         & President, Minneapolis   & Tr             & 2016 & $1{,}144$ \\
$11$ & Thomas I. Barkin      & President, Richmond      & Tr             & 2018 & $548$ \\
$12$ & Lorie K. Logan        & President, Dallas        & Tr, Bank       & 2022 & $606$ \\
$13$ & Susan M. Collins      & President, Boston        & Tr, Bank       & 2022 & $525$ \\
$14$ & Austan D. Goolsbee    & President, Chicago       & Tr, Bank       & 2023 & $249$ \\
$15$ & Beth M. Hammack       & President, Cleveland     & Tr, Bank       & 2024 & $180$ \\
$16$ & Jeffrey R. Schmid     & President, Kansas City   & Tr, Bank       & 2023 & $83$ \\
$17$ & Anna Paulson          & President, Philadelphia  & Tr, Bank       & 2025 & $145$ \\
\bottomrule
\end{tabular}
\caption{Onboarded roster. Channels: Sp~=~Board speeches, Te~=~testimony, Tr~=~FOMC transcripts,
PC~=~press conferences, BIS~=~BIS speech archive, Bank~=~home-bank speeches. ``From'' is the
collection start date, not necessarily the appointment date; ``Chunks'' is the per-member corpus size. ``ID'' is the member id used in Figure~1b of the main paper. Three presidents (Daly, Kashkari, Barkin)
are currently transcript-only; their home-bank speech scrapers are a staged follow-up.}
\label{tab:roster}
\end{table}

\paragraph{Members not onboarded.}
Two of the nineteen sitting members are absent. Alberto Musalem (St.\ Louis) publishes his
presidential remarks only on his home-bank domain, which refuses automated access at the network
level. We confirmed this block is robust: the host is unreachable from the collection
infrastructure (connection refused), from a server-side fetch service (HTTP~403), and from a
third-party Federal Reserve mirror (connection refused), while every other Federal Reserve and
aggregator host returns normally. The only externally reachable Musalem speeches predate his
presidency (2015--2016, from his tenure as a New York Fed executive) and were judged too stale and
off-role to represent his current monetary-policy stance, so he is excluded rather than onboarded
with misleading data. Cheryl Venable (Atlanta, interim) has no published speech or transcript
record. Both gaps are external-availability limits, not methodological ones.

\paragraph{Coverage boundaries.}
Beyond the two absent members, the collected record has principled boundaries that are worth
stating plainly, as they bound what each twin can represent. (i)~\emph{The 2011--2017 historical
gap is closed:} the back-fill collects the longest-serving members --- a governor from 2006, the
New York president from 2011, the chair from 2012, and Minneapolis from 2016 --- so aggregate
coverage is continuous from 2006 to the present (Figure~1 of the main paper); recent appointees simply
have no pre-tenure Federal Reserve record, which is a property of the institution, not the
collection. (ii)~\emph{Speech channels are wired per
office.} Governors come through the Board feeds; the New York president through BIS; six recently
appointed presidents through their home-bank pages. Three presidents (Daly, Kashkari, Barkin) have
no wired speech channel, since the BIS archive returns no speeches for them. They are therefore
represented only by their FOMC transcript turns, which, under the five-year embargo, currently end
in 2020. The underlying speeches are public; direct-site scrapers for these members are a clearly
scoped follow-up, not a limit of the public record. We report this because it bounds the recent
coverage of those twins and, in turn, their weight in the post-2020 index.

\section{Methods and Additional Analyses}
\label{app:methods}

\subsection{Models, embeddings, and metrics}
Each persona is $f_{\mathrm{base}}$ (\texttt{gpt-4o-mini}, temperature $0.2$) prompted with a
member-specific \texttt{FED\_MEMBER} biography and top-$k$ retrieval over the member's own chunks. All embeddings are \texttt{text-embedding-3-large} at $1024$ dimensions. Identifiability uses
a shrinkage Fisher LDA for $17$-way leave-one-query-out attribution on the shared queries
(\S5 of the main paper); LDA's pooled within-class covariance absorbs the shared per-query boilerplate. Detectability $\hat\tau_{\mathrm{det}}$ is a
cross-validated shrinkage Fisher LDA on generated-vs-real continuations, each pair truncated to
equal word length; $\hat\tau_{\mathrm{det}} = 2\max(A, 1{-}A) - 1$ for balanced accuracy $A$, with a real--real
floor. The hawk--dove axis is $\mathrm{normalize}(\overline{\mathrm{emb}}(\text{hawk anchors}) -
\overline{\mathrm{emb}}(\text{dove anchors}))$, fixed and model-independent; we use a single fixed
linear projection rather than a per-condition data-kernel embedding so conditions remain comparable.

\subsection{Identifiability: attribution confusion}
\begin{figure}[h]
\centering
\includegraphics[width=0.74\columnwidth]{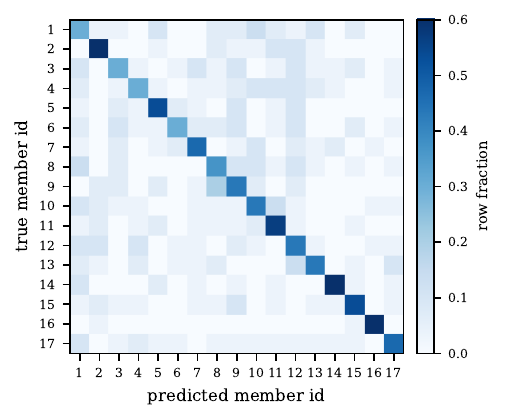}
\caption{Identifiability attribution confusion ($17$-way leave-one-query-out, shrinkage Fisher LDA on
raw responses, $k{=}3$), row-normalized so the diagonal is per-member recall. Member ids as in
Table~\ref{tab:roster}.}
\label{fig:confusion}
\end{figure}

Figure~\ref{fig:confusion} is the full $17$-way leave-one-query-out attribution at $k{=}3$ (shrinkage
Fisher LDA on raw responses, row-normalized so the diagonal is per-member recall; member ids as in
Table~\ref{tab:roster}). The diagonal dominates --- every member clears chance --- and the
off-diagonal mass is diffuse rather than blocky: misattributions spread across the committee rather
than concentrating on a single look-alike, so the residual confusion reflects shared committee
language, not indistinguishable pairs. The chair and centrist governors carry the lightest diagonals
(recall $\approx0.30$), consistent with their role as consensus voices.

\subsection{Detectability: \% of message presented}
The seed fraction (how much of the held-out quote is revealed) sets how much the persona must
generate, and thus its detectability (Table~\ref{tab:seed}, \texttt{gpt-4o-mini}, $k{=}3$): a short
seed leaves more to generate and is more detectable, a long seed less; half is our operating point.
Neither likeness axis is predicted by corpus size --- the thinnest twin ($83$ chunks, Schmid) is the
most identifiable, and detectability is uncorrelated with size (Kendall's
$\tau(\text{size}, \hat\tau_{\mathrm{det}}) = +0.23$, $p{=}0.21$); the consensus voices (centrist
governors, the New York president) are the least distinctive but still clear chance.

\begin{table}[h]
\centering\small
\begin{tabular}{@{}lc@{}}
\toprule
\% revealed & $\hat\tau_{\mathrm{det}}$ \\
\midrule
10\% & 0.55 \\
25\% & 0.53 \\
50\% & 0.23 \\
75\% & 0.18 \\
90\% & 0.20 \\
\bottomrule
\end{tabular}
\caption{\% of the held-out message revealed as the seed vs.\ detectability ($\hat\tau_{\mathrm{det}}$,
\texttt{gpt-4o-mini}, $k{=}3$): a shorter seed leaves more to generate and is more detectable;
$50\%$ is our operating point.}
\label{tab:seed}
\end{table}

\subsection{Query set: count and quality}
\label{app:stance}
We ask both how many queries are needed and which kind carries the signal. Count is not the lever:
within the curated $30$-question set, the reputational correlation saturates by $\sim$15--20
questions (Figure~3e of the main paper). For query \emph{type}, we score a balanced pool of $72$
spectrum-conditioned questions ($12$ hawk--dove facets, the query generator) per facet
(Figure~3f of the main paper): per-facet discrimination ranges from $\tau = 0.15$ to $0.38$, with the
strongest facets probing disposition and priorities (weight on labor slack vs.\ wage inflation, on
financial-stability risk, on how restrictive to be) and the weakest probing timing and tactics
(recession-risk appetite, preemptive tightening, patience before cuts). Good queries ask what a member
values, not what the Committee should do next meeting.

\paragraph{The curated query set.}
The $30$ monetary-policy questions used throughout (\S5 of the main paper) are listed below.
{\footnotesize
\begin{enumerate}\setlength{\itemsep}{-1pt}
\item Do you believe that the Federal Reserve should prioritize keeping inflation low over achieving maximum employment?
\item Do you believe that pursuing maximum employment should take precedence over maintaining price stability in the current economic environment?
\item Do you believe that a temporary increase in inflation is acceptable if it leads to significant job growth?
\item Do you believe that the overall economic health is best served by stabilizing prices even if it means higher unemployment rates in the short term?
\item Do you believe that the trade-off between inflation and unemployment justifies a more aggressive monetary policy approach?
\item Do you believe the Federal Reserve should prioritize reducing inflation below 2 percent even at the risk of higher unemployment?
\item Do you believe monetary policy should remain accommodative to support economic growth despite inflation readings above 2 percent?
\item Do you believe the Federal Reserve has the tools to effectively combat rising inflation without significantly hindering recovery efforts?
\item Do you believe maintaining a tolerable inflation rate above 2 percent could be beneficial for long-term economic stability?
\item Do you believe that current inflationary pressures warrant a preemptive increase in the federal funds rate to avoid long-term damage?
\item Do you believe the Federal Reserve should prioritize reducing inflation over maintaining low unemployment?
\item Do you believe that a temporary increase in unemployment is an acceptable trade-off for achieving long-term price stability?
\item Do you believe that the current inflationary environment justifies a more aggressive increase in the federal funds rate, even at the risk of higher unemployment?
\item Do you believe that the benefits of keeping inflation in check outweigh the social costs of rising unemployment rates?
\item Do you believe that the Federal Reserve should adopt a more dovish stance and focus on supporting employment despite persistent inflation?
\item Do you believe the Federal Reserve should raise interest rates now to preemptively combat potential inflationary pressures?
\item Do you believe that waiting to see actual signs of inflation before adjusting the federal funds rate is a prudent approach?
\item Do you believe that tightening monetary policy too early could stifle economic growth and recovery?
\item Do you believe that current economic indicators warrant a hawkish stance on interest rates to mitigate future inflation risks?
\item Do you believe that a dovish approach, focused on maintaining low rates, could be harmful if inflation expectations become embedded?
\item Do you believe that the current level of inflation justifies a more restrictive monetary policy approach?
\item Do you believe that maintaining a low federal funds rate is essential for promoting long-term economic growth?
\item Do you believe that rising wage pressures warrant a shift towards a more hawkish stance on interest rates?
\item Do you believe that the risks of an economic slowdown should caution the Federal Reserve against tightening policy too quickly?
\item Do you believe that addressing asset bubbles should take precedence over concerns about short-term inflation rates?
\item Do you believe that prioritizing financial stability over economic growth is essential for the long-term health of the economy?
\item Do you believe that maintaining a low federal funds rate is the best approach to stimulate growth in the current economic environment?
\item Do you believe that the risks of inflation should be weighed more heavily than the potential for financial instability when setting monetary policy?
\item Do you believe that a more aggressive tightening of monetary policy is necessary to prevent asset bubbles from forming in the current market?
\item Do you believe that the Federal Reserve should adopt a wait-and-see approach regarding interest rate changes to gauge the effects on both growth and financial stability?
\end{enumerate}
}

\paragraph{Hawk and dove anchors.}
The fixed dove$\to$hawk axis is $u=\widehat{\mathrm{hawk}}-\widehat{\mathrm{dove}}$, the difference of
the mean embeddings of the following hand-written anchor statements.
{\footnotesize
\emph{Hawkish:}
\begin{enumerate}\setlength{\itemsep}{-1pt}
\item Inflation is too high and we must raise interest rates decisively to restore price stability.
\item The risk of entrenched inflation outweighs concerns about slowing growth; policy should stay restrictive.
\item We should tighten monetary policy and shrink the balance sheet to combat inflationary pressures.
\item Premature easing would be a serious mistake; we must keep rates higher for longer.
\item Maintaining credibility on inflation requires a firm, hawkish stance even at the cost of some employment.
\end{enumerate}
\emph{Dovish:}
\begin{enumerate}\setlength{\itemsep}{-1pt}
\item We should cut interest rates to support employment and economic growth.
\item The labor market needs support; the risk of overtightening into a recession is significant.
\item With inflation falling we can afford to ease policy to protect jobs.
\item Accommodative monetary policy is warranted to sustain the recovery and maximize employment.
\item The greater danger now is doing too much and weakening the economy, so policy should be more dovish.
\end{enumerate}
}

\subsection{Generator robustness}
\label{app:generator}
\begin{figure}[h]
\centering
\includegraphics[width=0.6\columnwidth]{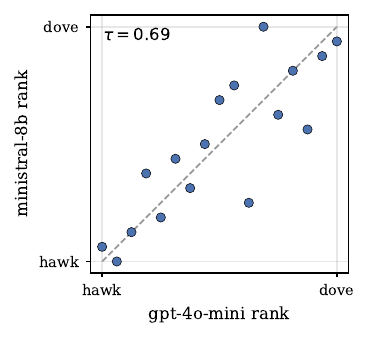}
\caption{Generator robustness: the persona-based member ordering ($k{=}3$) under two base models,
\texttt{gpt-4o-mini} (horizontal) vs.\ \texttt{ministral-8b} (vertical). Points on the dashed diagonal
agree (Kendall $\tau=0.69$).}
\label{fig:generator}
\end{figure}

The persona-based stance ordering is not an artifact of a single base model. Re-running the
persona-based estimate with \texttt{ministral-8b} in place of \texttt{gpt-4o-mini} yields a closely
agreeing member ordering across the $17$ personas (Kendall $\tau=0.69$, Figure~\ref{fig:generator}).

\subsection{Index: aggregation robustness}
The committee index is the mean of member axis-positions. It is robust to that choice
(Table~\ref{tab:agg}): median, trimmed-mean, and chair-weighted aggregations give
$\tau \in [0.64, 0.69]$ for the level$\leftrightarrow$move correlation, with the mean within noise
of the best on both metrics, so we keep it.

\begin{table}[h]
\centering\small
\begin{tabular}{@{}lcc@{}}
\toprule
Aggregator & $\tau$ (2022--25) & OOS 3-class acc \\
\midrule
mean & 0.68 & 0.69 \\
median & 0.64 & 0.62 \\
trimmed & \textbf{0.69} & 0.69 \\
chair $\times2$ & 0.68 & \textbf{0.72} \\
chair $\times3$ & 0.68 & 0.69 \\
\bottomrule
\end{tabular}
\caption{Index aggregation robustness (point-in-time corpus, pure-relevance retrieval). The simple
mean is within noise of the best on both metrics.}
\label{tab:agg}
\end{table}

\section{Limitations}
\label{app:limitations}
We note the boundaries of the present study. Detectability is read through one embedder
(\texttt{text-embedding-3-large}); a stronger or weaker discriminator would shift $\hat\tau_{\mathrm{det}}$, and we
validate indistinguishability statistically rather than with a human Turing test. Stance
is validated against a single, tenure-averaged reputational ordering (in the tradition of
\citealp{istrefi2019fedwatchers,bordo2023perceived}); a second source or a time-varying ground truth
would strengthen it, as would testing robustness to semantically-equivalent query paraphrases
\citep{helm2026generative} and to the choice of hawk--dove anchors. Hawk--dove is a single axis, and
the cut-vs-hold boundary suggests a second (``easing-urgency'') dimension. The index uses
latest-release rather than real-time (vintage) macro data, so the briefing $c^{(t)}$ can reflect
reference months whose official releases postdate the meeting by a few weeks. Residual generator
knowledge of realized outcomes is quantified rather than eliminated: on post-training-cutoff meetings
the walk-forward classifier matches but does not beat the hold-heavy base rate, while the level-lead
profile persists (\S6 of the main paper); the live index accumulates contamination-free meetings going
forward. Decision-relevant power concentrates in the well-covered $2022$--$2025$ window. These are
scoped limitations, not method failures, and each has a concrete next step.


\end{document}